\pdfoutput=1 

\documentclass[]{aa}
\usepackage{graphicx}
\usepackage{latexsym}
\usepackage{graphicx}
\usepackage{subfigure}
\usepackage{longtable}
\usepackage{supertabular}
\usepackage{natbib}
\usepackage[latin1]{inputenc}
\usepackage[thinspace]{SIunits}
\usepackage{graphicx}
\usepackage{graphics}
\usepackage{epsfig}

\begin{document}

\title{Large scale plane--mirroring in the cosmic microwave background WMAP5 maps}
\author{V.G.Gurzadyan\inst{1,2},  A.A.Starobinsky\inst{3},  T.Ghahramanyan \inst{1},  A.L.Kashin\inst{1}, \\
H.G.Khachatryan\inst{1},  H.Kuloghlian\inst{1}, D.Vetrugno\inst{4} and G.Yegorian\inst{1}}

\institute
{
\inst{1} Yerevan Physics Institute and Yerevan State University, Yerevan,
Armenia\\
\inst{2} ICRANet, ICRA, University ``La Sapienza'', Rome, Italy\\
\inst{3} Landau Institute for Theoretical Physics, Moscow, 119334, Russia\\
\inst{4} University of Lecce, Lecce, Italy
}

\date{Received (\today)}

\titlerunning{Large scale plane-mirroring in CMB maps}

\authorrunning{V.G.Gurzadyan et al.}

\abstract{We continue investigation of the hidden plane-mirror symmetry in the distribution of
excursion sets in cosmic microwave background (CMB) temperature anisotropy
maps, previously noticed in the three-year data of the Wilkinson microwave
anisotropy probe (WMAP), using the WMAP 5 years maps. The
symmetry is shown to have higher significance, $\chi^2 < 1.7$, for low
multipoles $\ell < 5$, while disappearing at larger multipoles, $\chi^2 >
3.5 $ for $\ell > 10$. The study of the sum and difference maps of
temperature inhomogeneity regions, along with simulated maps, confirms its
existence.The properties of these mirroring symmetries are compatible with
those produced by the Sachs-Wolfe effect in the presence of an anomalously
large component of horizon-size density perturbations, independent of one of
the spatial coordinates, and/or a slab-like spatial topology of the Universe.}


\maketitle

\section{Introduction}

The properties of the CMB temperature anisotropy, as well as its
polarization, are among the basic sources of information on cosmological
parameters \cite{DB1,Sp,Kom}. Their tiny features, such as the local spikes in the
multipoles power spectrum, deviation from the statistical isotropy, and
non-Gaussianity signatures, may all be the result of various fundamental
processes having occurred in the early Universe. Among the reported anomalies
are the alignment of the principal (Maxwellian) vectors of low multipoles,
the north-south power asymmetry, the southern anomalous cold spot, etc 
(see de Oliveira-Costa et al. 2004; Copi et al. 2004,2007; Schwarz et al. 2004;
Eriksen et al. 2004,2007; Cruz et al. 2005; Morales \& Saez 2008).
In the present paper
we continue the study of another deviation from statistical isotropy: the
hidden partial plane-mirror symmetry in the distribution of CMB temperature
fluctuations excursion sets, i.e. of one-connected pixel sets equal to and
higher than the given temperature threshold (lower for negative thresholds)
previously found in the WMAP 3-years temperature maps \cite{mirr}. We use
the WMAP 5-years data \cite{WMAP5}, not only to confirm the mirroring effect
found in WMAP3 maps while studying the inhomogeneities in the distribution
of the excursion sets, but also to reveal its other properties. By
inquiring into the dependence of the mirror symmetry on the angular scale,
we show that the effect has the highest significance at low multipoles $\ell
<5$ and that it quickly disappears at higher multipoles.

Similarly, the study of the sum and difference maps of temperature inhomogeneity
regions provides additional insight into the mirroring. Namely, when the sum
and difference maps are created via reflection of one of the maps, as it
should be for mirrored images, anisotropic properties of excursion sets do
survive, while they disappear if the sum map is created without reflection.
Difference maps from independent radiometers (A-B) have been used to test
the role of scan inhomogeneities and noise \cite{mirr}. The signal-to-noise
ratio for the studied excursion sets is about 4:1 and the excursion sets in
(A-B) map do not show any specific property observed in the sum (A+B) map.
The negligible role of the noise was also checked using the foreground reduced
maps available in http://lambda.gsfc.nasa.gov/product/map/current/.
Although contamination of a Galactic or interplanetary origin at these
multipoles certainly cannot be excluded, following Gurzadyan et al (2007b), in the last
section we discuss which properties of the Universe might be responsible for
this effect, if it had a cosmological origin.

\section{Distribution of excursion sets}

For this analysis we used the 94 GHz (3.2mm) W-band WMAP 5-year maps, due to
their offering the highest angular resolution (beam width of FWHM=$0^{\circ }.21$) and
lowest contamination by synchrotron radiation of the Galaxy \cite{Ben}. 
The role of the Galactic disk was minimized via exclusion of the
equatorial belt $|b|<20^{\circ }$.

Algorithms for studying the excursion sets have been described in \cite%
{Gur1a,Gur2} in connection with the study of the ellipticity in excursion
sets in the Boomerang and WMAP, and earlier for COBE maps \cite{GT}. The definition
of the centers and of the geometrical characteristics of the excursion sets
are based on rigorous procedures, e.g. the Cartan theorem for the
conjugation of the maximally compact subgroups of Lie groups. The distribution
of the centers of the excursion sets obtained via those algorithms were
obtained for various pixel count and temperature-threshold interval sets and the 
center of these $N$ centers was obtained: $x$(center)=$\sum x_i/N$, $y$(center)=$\sum y_i/N$.  

Inhomogeneities in the distribution of the excursion sets at the temperature
interval $\Delta T =90~\mu K$ within $|T|=45~\mu K$ are concentrated around
almost antipodal points centered at 
\begin{eqnarray}
l&=& 94^\circ.7,\,\,\,\, b= 34^\circ.4\,\, (CE_N);  \nonumber \\
l&=& 279^\circ.8,\,\,b= -29^\circ.2\,\, (CE_S).  \nonumber
\end{eqnarray}
At higher temperature interval limits the excursion sets are merged and
loose their identity. The methodical novelty here is 
that we consider the excursion sets within a 
temperature interval $[-T,T]$ with respect to
the mean CMB temperature, instead of excursion sets
in the slice around a fixed $T$. Namely, {\it the symmetry appears
at the sum of slices but not in each of them}. This is analogous to the appearance
of the Great Wall as a result of the sum of slices of large scale galaxy surveys.    

Comparing these positions to those of the Maxwellian vectors of the lowest
multipoles of CMB, it was shown that $CE_N$ and $CE_S$ are located close to
one of the vectors of multipole $\ell =3$, shifting towards the equator when
increasing the temperature interval. During this shift, the mirror symmetry
is approximately maintained, while the patterns of the excursion sets around 
$CE_N$ and $CE_S$ mirror each other with $\chi^2 = 0.7 - 1.5$.

Neither $CE_N$ and $CE_S$ are close to the positions of the sum of the
multipoles vectors with $\ell=2-8$, the modulus of each vector weighted by $%
1/\ell(\ell +1)$). The position of $CE_S$ is not close to that of the
cold spot \cite{Vielva}.

\section{Mirroring versus multipoles}

We now investigate the multipole dependence of the mirroring. We study it
using the method of gradually removing multipoles, defined by the
coefficients $a_{\ell m}$ of the temperature fluctuations expansion into
spherical harmonics: 
\begin{equation}
\frac{\Delta T_{\ell }(\hat{n})}{T}=\sum_{m=-\ell }^{\ell }a_{\ell m}Y_{\ell
m}(\hat{n}).
\end{equation}
The functions ``anafast" and ``synfast" of Healpix \cite{Healpix} were used,
and the WMAP5 94 GHz (W channel) dataset was analyzed. We checked that the
temperature anisotropy can be represented as 
\begin{equation}
\frac{\Delta T(\theta, \phi)}{T}=\left(\frac{\Delta T}{T}\right)_{mirr}+
\left(\frac{\Delta T}{T}\right)_{non-mirr}  \label{split}
\end{equation}
where the first, mirrored term dominates at certain (low) multipoles, while
the second, non-mirrored term becomes the main one at other (higher)
multipoles. Such partial mirroring does not imply planarity, i.e.
the dominance of multipoles with $|m|=\ell$. Instead, for the first term on
the right hand side of Eq. (\ref{split}), all $a_{\ell m}$ 
with an even value of $\ell - m$ may be non-zero.

To quantify the degree of mirroring in \cite{mirr}, the drift of the centers of the
symmetries with respect to the dipole apexes vs the temperatures for each
hemisphere was studied. The confidence levels defined as $\chi ^{2}=
(\sum_{i}^{N}(Y_{1i}-Y_{2i})^{2})/(N-DOF)$ (where $Y_{1i}, Y_{2i}$ are the distances to the
northern and southern dipole apexes, respectively, and $N\simeq 30$ is the number of steps,
$DOF=2$, for the compared curves) have been
evaluated, i.e. the departure from being identical (see
Fig. 4 in \cite{mirr}, as well as Figs. 5 and 6 there for the distances to the
the multipole vectors). Here we evaluate the analogeous confidence
levels but for the maps split over multipoles. 
Figure 1 shows the plots for the excursion sets of more than 50 pixel counts vs the
multipole numbers where the points of given $\ell$ indicate that multipoles
smaller than $\ell$ are cancelled. The red line there represents 
the best polynomial fit, 
while the dashed
lines correspond to smoothed error bars. The plot shows that the mirroring
effect has its highest significance at low multipoles, $\chi ^{2}<1.7$ for $%
\ell <5$, and it weakens monotonically for higher ones; i.e. $\chi
^{2}>3.5$ at $\ell >10$. For comparison, we also show the same dependence
obtained using the WMAP 3-year W-maps (Fig. 2). Similar dependence occurs
for larger excursion sets (Fig. 3). Figure 4 shows confidence levels for the
mirroring effect when one of the mirrored regions (the northern, see below)
is replaced by a simulated, statistically isotropic Gaussian map. For the latter,
first, the maps for the given multipoles were obtained from the real map, then,
for the resulted multipole maps, isotropic Gaussian maps were generated using the mean
and the $\sigma$ of each map.    

Then, to probe the mirroring further, we constructed and compared the sum and
the difference maps of regions ($l,b$: $[10^{\circ},170^{\circ}]$, $%
[20^{\circ},90^{\circ}]$; $[190^{\circ},350^{\circ}]$, $[-20^{\circ},-90^{%
\circ}]$), respectively centered on $CE_N$ and $CE_S$, using two procedures:
(a) with rotation on the angle $\pi$, i.e. keeping the mirror symmetry; (b)
without this rotation. The number of the excursion sets surrounding $CE_N$ and $CE_S$
are given in Fig. 5. The angular distances of $CE_N$ and $CE_S$ from the
CMB dipole direction vs the temperature threshold are shown in Figs. 6a and
5b for the mirrored (i.e. with rotation) and non-mirrored sums and difference
maps, respectively; the continuous and dashed lines denote the sum and
difference maps, respectively. In Fig. 6b, the same dependence is plotted
for a simulated statistically isotropic map, too (the upper dot-bar curve).
Note that, while a difference map (A-B) obtained using data from different
radiometers but from the \textit{same} region of the sky contains mainly
noise and no signal, when we deal with \textit{different} sky regions, the
difference map should result in another map that is not very different from the sum
map. Indeed, Figs. 6a and b clearly indicate that the similarity in the
temperature independence of the distance from the dipole for the
non-mirrored sum map and the simulated map is obvious -- in both maps there
is no breaking of statistical isotropy. This differs crucially from the
temperature dependence of this distance in the case of the mirrored sum and
difference maps (Fig. 6a), thus confirming the existence of a partial mirror
symmetry of the regions of $CE_N$ and $CE_S$.

\begin{figure}[ht]
\centerline{\epsfig{file=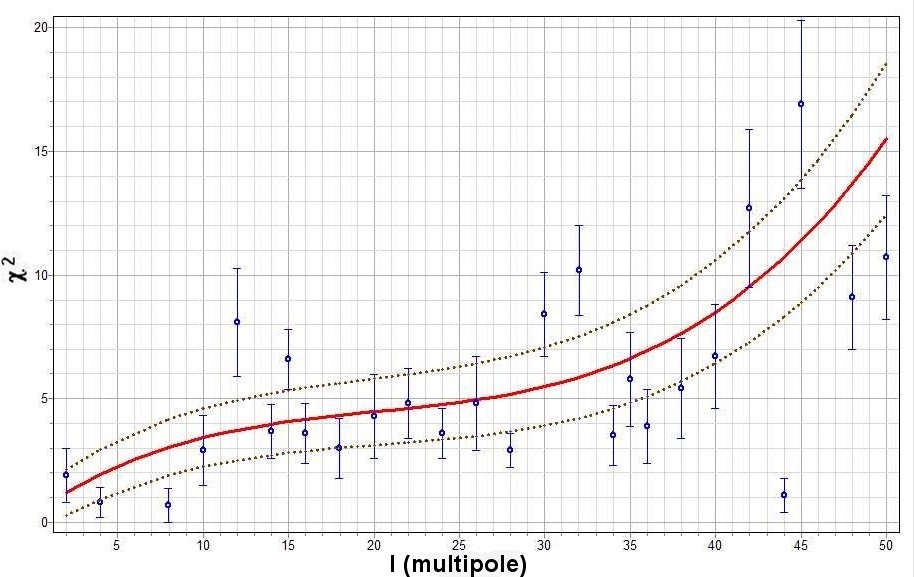,width=0.5\textwidth}} \vspace*{8pt}
\caption{Statistical significance of the mirroring of patterns of excursion
sets with more than 50 pixel counts around the centers $CE_N$ and $CE_S$ vs
the multipole number $\ell$ for the WMAP 5-year 94 GHz temperature maps.}
\end{figure}

\begin{figure}[bp]
\centerline{\epsfig{file=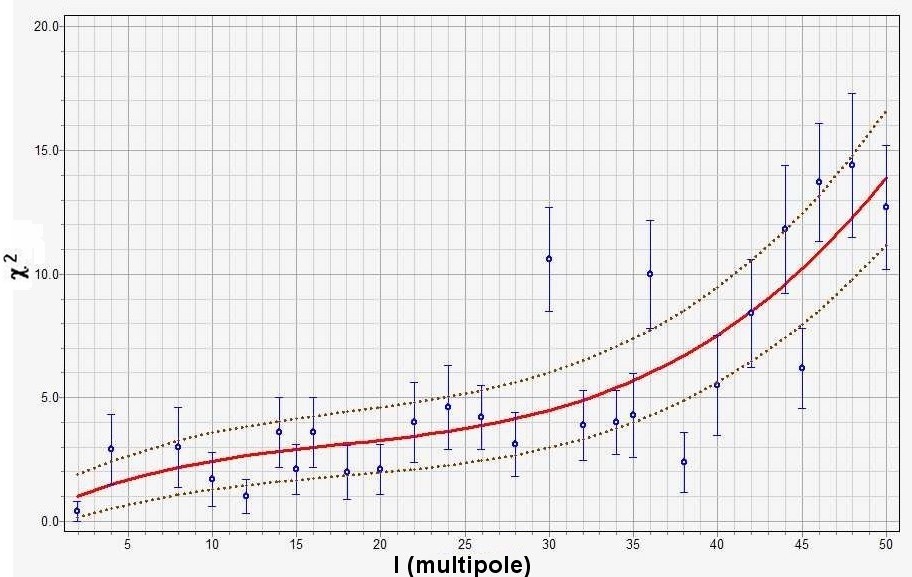,width=0.5\textwidth}} \vspace*{8pt}
\caption{The same as in Fig. 1, but for the WMAP 3-year maps.}
\end{figure}

\begin{figure}[ht]
\centerline{\epsfig{file=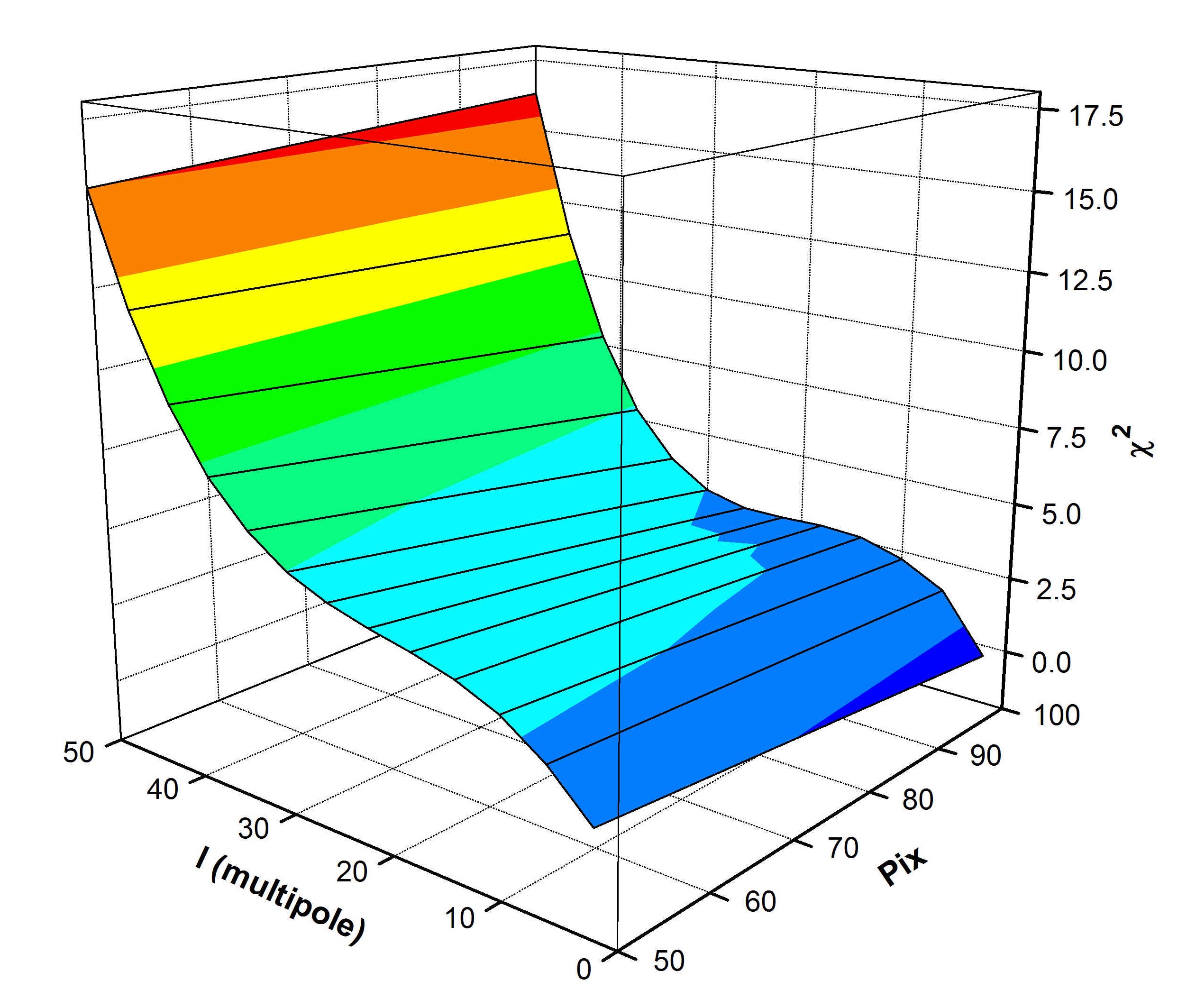,width=0.5\textwidth}} \vspace*{8pt}
\caption{$\protect\chi^2$ dependence as in Fig. 1, but now both vs the
multipole number and the number of pixel counts of the excursion sets.}
\end{figure}

\begin{figure}[tp]
\centerline{\epsfig{file=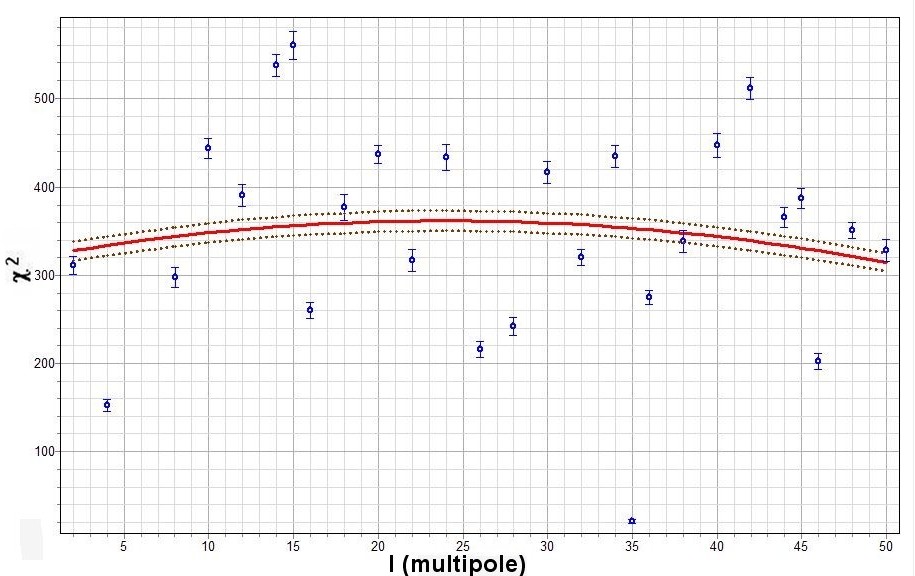,width=0.5\textwidth}} \vspace*{8pt}
\caption{$\protect\chi^2$ when one of the symmetry regions is replaced by a
simulated statistically isotropic map.}
\end{figure}

\begin{figure}[bp]
\centerline{\epsfig{file=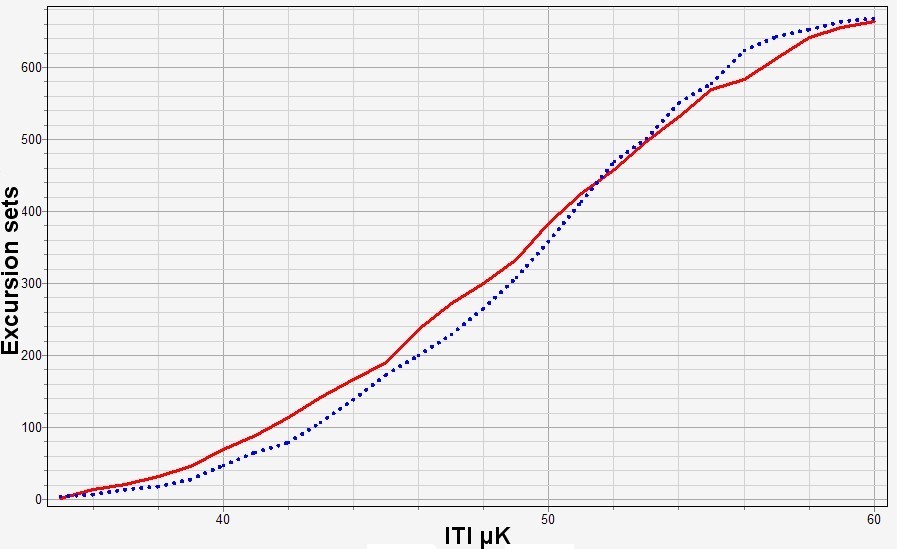,width=0.5\textwidth}} \vspace*{8pt}
\caption{The number of excursion sets with more than 100 pixels vs 
the temperature threshold within the regions centered on $CE_N$ and $CE_S$ (dashed line)
for the WMAP 5-year 94 GHz temperature maps.}
\end{figure}

\begin{figure}[tp]
\centerline{\epsfig{file=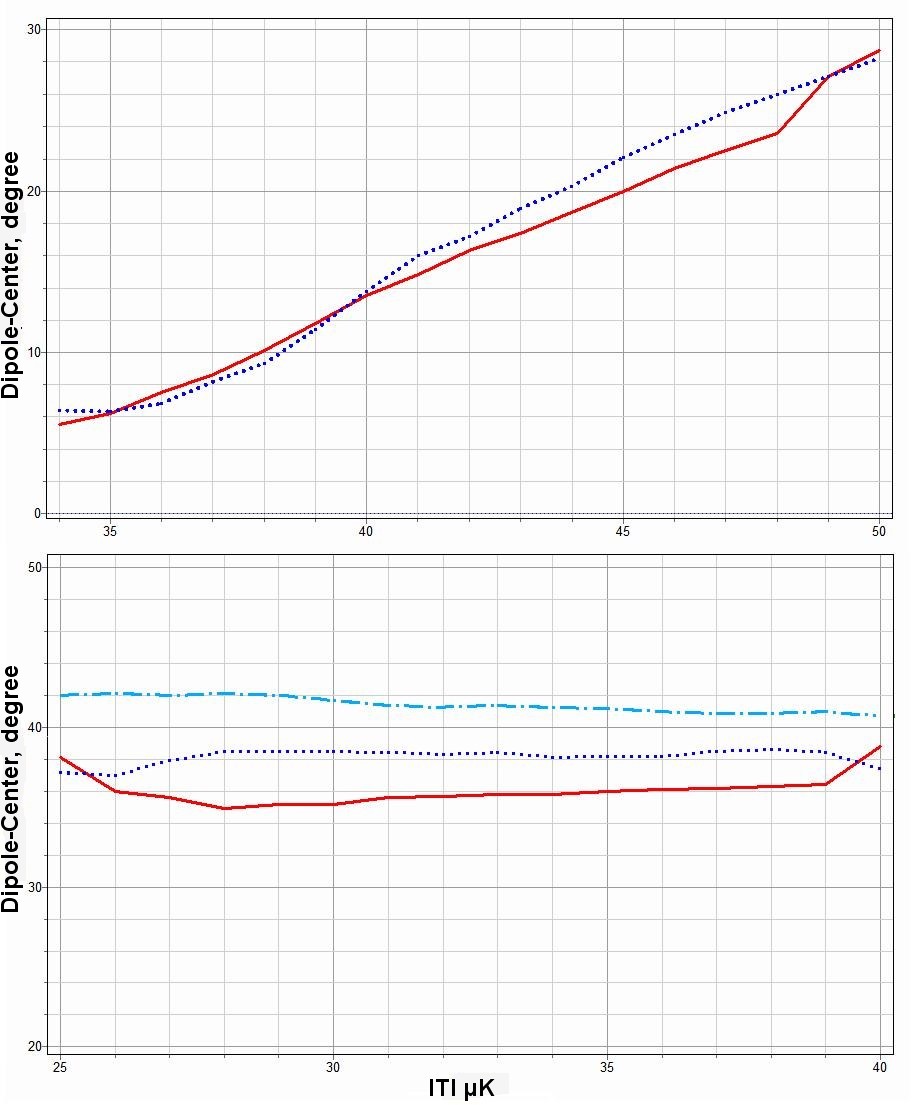,width=0.5\textwidth}} \vspace*{8pt}
\caption{Distances of $CE_N$ and $CE_S$ from the CMB dipole vs temperature
for the sum and difference maps centered on $CE_N$ and $CE_S$. (a) The sum
and difference (dashed) maps obtained via rotation by the angle $\protect\pi$
in one of the maps (to keep the mirror symmetry); (b) no mirroring rotation
performed for the sum and difference (dashed) maps; for comparison, the case
of simulated isotropic maps is shown (dot-bar curve).}
\end{figure}
\section{Conclusions and discussion}

We have analyzed the scale dependence of partial mirror symmetry in the
distribution of excursion sets, as previously found in the WMAP3 maps, with a
nearly antipodal location of symmetry centers. Studies 
used the WMAP5 W-band maps, and the results obtained using WMAP5 and WMAP3 data
agree. The centers lie close to one of the $\ell=3$ multipole Maxwellian
vectors, but not close to the sum of multipoles vectors up to $\ell=8$ \cite%
{mirr}. Also they are close to the ecliptic pole and are nearly orthogonal
to the CMB dipole apexes. They are moving towards the Galactic equator with
the increase in the temperature threshold interval.

This symmetry appears to be a large-angle effect, i.e. it is stronger at low
multipoles and it weakens rapidly  for larger $\ell$: its statistical
significance is quantified by $\chi^2 < 1.7$ at $\ell < 5$ and $\chi^2 > 3.5$
at $\ell > 10$. The symmetry was also tested using the
following procedure: the sum map of the symmetry regions was obtained
first, via rotation over $\pi$, as is usually the case for mirrored images, and then
without such a rotation. The clear mirroring in the first case and its
complete absence in the second case make the case for a partial mirror
symmetry stronger.

Turning to the origin of this symmetry, an unknown interplay of
interplanetary and Galactic foregrounds or another unspecified
non-cosmological contribution to the low multipoles certainly remains a
possibility. If, however, it has a cosmological origin, then a signature of
the simplest non trivial, $T^1$, spatial topology of the Universe is among
the options, as discussed in Gurzadyan et al (2007b). For this topology, the points with
coordinates $z$ and $z+L$ are identified where $z$ is one of the spatial
coordinates. Such a model may be also considered as a limiting case of the
Universe with compact flat spatial sections having the $T^3$ topology if the
identification scales $L_1,L_2$ along two other spatial coordinates are much
more than $L$ -- the slab topology (for early papers on a non trivial
spatial topology of the Universe, see Zeldovich (1973), Sokolov \& Schwartsman (1975),
Sokolov \& Starobinsky (1975), Fang \& Sato (1983)). 
Note that, one should not expect any mirror symmetry, even a partial one, for
comparable topological scales $L\sim L_2,L_3$.

As shown in Starobinsky (1993), for this $T^1$ topology, a large-angle pattern
of a CMB temperature anisotropy has just the form (\ref{split}). The first
term on its right hand side has the exact mirror symmetry with respect to
the $(x,y)$-plane. It originates from the Sachs-Wolfe effect at the last
scattering surface from density perturbations that do not depend on $z$.
The second term represents a remaining part of anisotropy and does not have
any symmetry at all. However, for $a_0L$ on the order of $R_{\mathrm{hor}}$
or slightly more, where $a_0=a(t_0)$ is the present scale factor of a
Friedmann-Robertson-Walker cosmological model, the latter term should 
somehow be suppressed since the Sachs-Wolfe contribution to it from the last
scattering surface comes from perturbations having wave vectors with $|%
\mathbf{k}|\ge 2\pi/L$. That is why one expects the total large-angle
pattern of $\Delta T/T$ to have an \emph{approximate} mirror symmetry in
this case. \footnote{%
Note the other effect worsening the symmetry at large angles (low
multipoles): a contribution from the integrated Sachs-Wolfe effect at small
redshifts due to a cosmological constant (first calculated in \cite{KS85})
or dynamical dark energy.}

More generically and not connected with a non trivial spatial topology of
the Universe, such an approximate mirror symmetry at large angles arises
when the large-scale $z$-independent part of density perturbations inside
the last scattering surface is anomalously large. In all these cases, the
rms amplitude of the first term on the right hand side of Eq. (\ref{split})
quickly becomes negligible compared to the second one with the growth of $%
\ell$. Weakening of the mirroring symmetry for higher values of $\ell$ found
in Sect. 3 is in a good agreement with this theoretical prediction and may be
considered as an additional argument for the reality of the mirroring effect.

Until now, searches for the mirroring effect of the form (\ref{split}) or
directly for the $T^1$ topology gave negative results (see e.g.
de Oliveira-Costa et al. 1996; de Oliveira-Costa et al. 2004; Cornish et al. 2004), 
rasing the lower limit on the physical topological scale 
$a_0L$ to $\sim R_{\mathrm{hor}}=14$ Gpc. The numerical value is given
for the standard $\Lambda$CDM cosmological model with $\Omega_m=0.3,~%
\Omega_{\Lambda}=0.7$. However, for higher values of $a_0L$ this topology
is not excluded.
This follows already from the fact that the much more restrictive 
cubic $T^3$ topology ($L=L_1=L_2$) with $a_0L > R_{hor}$ is still 
considered as a viable possibility; see the recent papers \cite{AJLS07,A08}, 
where an inconclusive evidence for the 
latter case with $a_0L \approx 1.15R_{hor}$ is presented.
This shows that topological
explanation of the partial mirror symmetry investigated in this paper is
possible. From our analysis, it is still too early to speak about the
value of $L$, since the secure separation of CMB temperature fluctuations
into a mirrored and non-mirrored parts needs higher resolution maps. Also, a
non-topological (though still cosmological) explanation of such an effect is
possible, as pointed out above. In this respect, see the recent papers 
\cite{GK2008} where it was shown that voids can act as hyperbolic lenses in
a spatially flat Universe, producing specific signatures in CMB temperature
fluctuations. Future observational data will help solve these problems.

Shortly before submission of this manuscript, a paper appeared \cite{GE}
where CMB statistical anisotropy of an axial type was studied with the
preferred axis very close to the one defined by our $CE_N-CE_S$ direction.

\section{Acknowledgments}

We thank the referee for valuable comments. 
We are grateful to Paolo de Bernardis for continuous advice and help. 
AAS
was partially supported by the Research Program ``Astronomy" of the Russian
Academy of Sciences and by grant LSS-4899.2008.2. YPI team was partially
supported by INTAS. HKh was supported by the IRAP PhD program.

\end{document}